\documentclass[a4paper,11pt]{article}
\usepackage{geometry}
\usepackage[english]{babel}
\usepackage[utf8]{inputenc}
\usepackage[T1]{fontenc}
\usepackage{indentfirst}
\usepackage{amsmath}
\usepackage{amssymb}
\usepackage{eufrak}
\usepackage{graphicx}
\usepackage{psfrag}

\allowhyphens

\newcommand{\complex}{\mathbb{C}}
\newcommand{\fa}{\mathfrak{a}}

\newcommand{\valos}{\mathbb{R}}

\newcommand{\ordo}{\mathcal{O}}

\newcommand{\ket}[1]{{\left|#1\right\rangle}}
\newcommand{\bra}[1]{{\left\langle #1\right|}}

\newcommand{\skalarszorzat}[2]{{\langle #1 | #2 \rangle}}


\usepackage{ifpdf}

\ifpdf
\usepackage{epstopdf}
\usepackage[pdftex,colorlinks,urlcolor=blue,citecolor=blue,linkcolor=blue]{hyperref}
\else
\usepackage[hypertex,colorlinks,urlcolor=blue,citecolor=blue,linkcolor=blue]{hyperref}
\fi
\begin{document}
\numberwithin{equation}{section}

\title{Overlaps with arbitrary two-site states in the XXZ spin chain}
\author{B. Pozsgay$^{1,2}$\\
~\\
  $^1$ Department of Theoretical Physics, Budapest University\\
	of Technology and Economics, 1111 Budapest, Budafoki \'{u}t 8, Hungary\\
$^2$ BME Statistical Field Theory Research Group, Institute of Physics,\\
	Budapest University of Technology and Economics, H-1111 Budapest, Hungary}

\maketitle

\abstract{We present a conjectured exact formula for overlaps between
the  Bethe states of the spin-1/2 XXZ chain and generic two-site
states. The result takes the same form as in the previously known
cases: it involves the same ratio of two Gaudin-like determinants,
and a product of single-particle overlap functions, which can be fixed
using a combination of the Quench Action and Quantum Transfer 
Matrix methods. Our conjecture is confirmed by numerical data from
exact diagonalization.
For one-site states the formula is found to be correct even in chains
with odd length, where existing methods can not be applied.
It is also pointed out, that the ratio of the Gaudin-like
determinants plays a crucial role in the overlap sum rule:
it guarantees that in the thermodynamic limit there remains no $\mathcal{O}(1)$ piece in the Quench Action.
}

\section{Introduction}

Integrable quantum mechanical models are special many-body theories,
where the eigenvalues and eigenstates of the Hamiltonian are known
exactly \cite{korepin-book,sutherland-book}. Moreover, in many cases even
thermodynamical properties and correlation functions can be computed
\cite{Takahashi-book,kluemper-goehmann-finiteT-review,karol-hab}. There
is a vast literature devoted to the study of 
equilibrium properties of models such as the XXX and XXZ chains, which
play a central role due to their relative simplicity despite their
interacting nature, and their experimental relevance. Recently
an increasing attention has been focused on the study of overlaps. The
main question is: how can we compute the overlaps between exact
eigenstates of integrable models and certain ``initial states'', and
what is the class of states for which exact analytic results are to be
expected. The motivation to study this problem comes from (at least)
three different areas of research: Quantum Quenches, the AdS/CFT
conjecture, and finite temperature problems. 

First of all, the study of non-equilibrium 
situations raised the question whether isolated integrable systems
equilibrate to Generalized Gibbs Ensembles (GGE's)
\cite{rigol-gge,essler-fagotti-quench-review,rigol-quench-review}. One
way towards a definite answer is to compute the long time limit of
observables from first principles and to compare it to the prediction
of the GGE. This approach relies on the knowledge of the overlaps:
they are used as an input for the Quench Action (QA) method
\cite{quench-action}, which selects those Bethe states that are
relevant for the long-time limit of local observables. Thus it was
necessary to obtain exact formulas for the overlaps with Bethe states,
and this program has been carried out for certain initial states of
the spin-1/2 XXZ model
\cite{sajat-neel,Caux-Neel-overlap1,Caux-Neel-overlap2,Brockmann-BEC}. These
results lead to the understanding that the GGE built from the
so-called ultra-local charges is not sufficient to describe the
stationary states \cite{JS-oTBA,sajat-oTBA}, but an extended GGE that
incorporates all the (recently discovered) quasi-local charges
\cite{enej-review} gives
correct predictions \cite{JS-CGGE}, at least for the XXZ chain.

The Complete GGE specified in \cite{JS-CGGE} (see also
\cite{doyon-gge,enej-gge,sajat-eric}) is self-sufficient without 
the knowledge of the overlaps: it completely determines the relevant Bethe
states through certain relations called the ``string-charge duality''
\cite{jacopo-massless-1}. The effectiveness of this approach  has been
demonstrated for different types of initial states in \cite{JS-CGGE,eric-lorenzo-exact-solutions,sajat-eric}.
This raises the question whether the study of overlaps is still
relevant for the non-equilibrium problems. We believe the answer is a
definite yes. On the one hand, the overlaps are necessary ingredients for certain methods
 computing the finite time dynamics
 \cite{finite-qa,vincenzo-QA,vincenzo-renyi2}. On the other hand,
 in systems with higher rank symmetries the overlaps ``lead the way''
 once again: there are cases where the QA method is worked out \cite{nested-quench-1}
 using exact overlap formulas \cite{ADSMPS2}, but the GGE is not yet
established.
Also, a different viewpoint was laid out in
 \cite{sajat-integrable-quenches}, which focused on the question of
 which initial states can be considered ``integrable'', i.e. when do
 common features of integrability show up in the overlaps, the time evolution, and the stationary states. It was
 argued in  \cite{sajat-integrable-quenches} that very generally it
 is the subclass of integrable initial states where exact (factorized)
 overlap formulas can be expected. Having a precise definition of
 integrability and different methods to test it, it is now an
 interesting problem to find exact overlap formulas for integrable
 initial states, beyond the known cases, both in the spin-1/2 XXZ
 chain and other models. Closely related questions
 about overlaps have been
 investigated in Integrable QFT's in the papers
 \cite{gabor-boundary-state1,Gabor-initial-states-QFT-quench,gabor-quench-overlaps}.

A second motivation to study exact overlap formulas comes from the
AdS/CFT conjecture, where the overlaps describe one-point functions of
composite operators
\cite{zarembo-neel-cite1,zarembo-neel-cite3}. This line of
research led to new overlap formulas, which include overlaps with
integrable MPS's in the spin-1/2 XXX chain
\cite{zarembo-neel-cite1,zarembo-neel-cite2,zarembo-neel-cite3} and
the $SU(3)$-symmetric model too \cite{ADSMPS2}. In the XXX
chain the MPS's in question were shown to be zero-momentum
components of two-site states or finitely entangled states obtained by
the action of transfer
matrices on two-site states, whereas the interpretation of the MPS of
\cite{ADSMPS2} within integrability is not yet known. 

Finally, the third motivation to study overlaps comes from finite
temperature problems: It was shown in \cite{sajat-karol} that the
boundary free energy of the XXZ spin chain can be calculated in the
Quantum Transfer Matrix (QTM) framework by the Trotter limit of an exact
overlap (see also \cite{Goehmann-Bortz-Frahm-boundaries}). The
situation is analogous to integrable boundary QFT, where the object
in question is the exact overlap between the finite volume ground
state and a boundary state \cite{Dorey:1999cj,Dorey:2009vg}, although
in QFT this quantity is computed more easily from Thermodynamic Bethe
Ansatz \cite{sajat-g,woynarovich-uj}.

Regarding the XXZ spin chain, existing overlap formulas
\cite{sajat-neel,Caux-Neel-overlap1,Caux-Neel-overlap2,Brockmann-BEC,zarembo-neel-cite2} concern
two-site states that are described by the diagonal $K$-matrices (see
main text for definitions). The most studied examples are the N\'eel
and dimer states. On the other hand, there are exact results available for
quenches from other two-site states \cite{sajat-Loschm}, although
these results apply to the thermodynamic limit directly. It was shown in \cite{sajat-integrable-quenches}
that in the spin-1/2 case all two-site states are integrable,
therefore it is natural to expect that there are exact finite volume
overlap formulas for arbitrary two-site states. The construction of
such formulas is the problem that
we investigate in the present work.

In the next section we collect a number of existing
results for overlaps and quenches, and we point out an important
link between the relevant finite volume and infinite volume
calculations. This connection is then used in Section \ref{sec:ov} to
conjecture an exact overlap formula for generic $K$-matrices. At
present we do not have a proof of our result; numerical checks
confirm its validity, and it is shown to have the correct
thermodynamic limit. In Subsection \ref{sec:ovsumrule} we also point
out an interesting relation between the Gaudin-like determinants
appearing in the overlaps and the calculation of the overlap sum rule
within the Quench Action framework.

\section{Overlap formulas -- Ingredients}

\label{sec:ingred}

\subsection{The model and its Bethe Ansatz solution}

We consider the anti-ferromagnetic spin-$1/2$ XXZ Heisenberg model on a chain of length
$L$ with periodic boundary conditions. The Hamiltonian is
\begin{equation}
  \label{XXZ-H}
  H=\sum_{j=1}^{L}
  (\sigma^x_j\sigma^x_{j+1}+\sigma^y_j\sigma^y_{j+1}+\Delta
(\sigma^z_j\sigma^z_{j+1}-1)).
\end{equation}
Here $\Delta$ is the anisotropy parameter. In this work  we will focus
on the massive  regime ($\Delta>1$), but we expect that our results
will be applicable to most states even in the $\Delta<1$
case\footnote{
  This follows from the fact that the finite volume overlap formulas
  arise from a set of algebraic manipulations on
  the Bethe Ansatz wave function, which has the same functional form
  for every $\Delta$. The differences between the regimes only show up
  at the solutions of the Bethe equations (including various types of
  singular rapidities \cite{baxter-completeness}), and in the
  thermodynamic limit. However, these issues of the massless regime are not considered here.
}.

This Hamiltonian can be diagonalized by the Bethe Ansatz
\cite{XXX,XXZ1,XXZ2,XXZ3}. The eigenstates are constructed as
interacting spin waves over a ferromagnetic reference state and they
are characterized by a set of rapidities $\{\lambda\}_N$; for the
states we will use the notation $\ket{\{\lambda\}_N}$.
In the coordinate Bethe Ansatz representation the 
un-normalized wave function can be written as
\begin{equation}
  \label{BA2}
 \Psi_{L}(\lambda_1,\dots,\lambda_N|s_1,\dots,s_N)=\sum_{P\in\sigma_N} 
\prod_j  F(\lambda_{P_j},s_j) 
\prod_{j>k} \frac{\sin(\lambda_{P_j}-\lambda_{P_k}-i\eta)}{\sin(\lambda_{P_j}-\lambda_{P_k})}
\end{equation}
with
\begin{equation}
\label{Fs}
  F(\lambda,s)=\sinh(\eta)
\sin^{s-1}(\lambda+i\eta/2)
\sin^{L-s}(\lambda-i\eta/2).
\end{equation}
Here $s_j$ denote the positions of the down spins over a ferromagnetic
reference
state with all spins up, and we assume
$s_j<s_k$ for $j<k$. The parameter
$\eta$ is given by the relation $\Delta=\cosh\eta$ and the rapidities
$\{\lambda_j\}$ characterize the spin waves. The Bethe wave function
is invariant with respect to a shift $\lambda_j\to\lambda_j+\pi$ for
every $j$ (up to an irrelevant phase
for odd $L$), therefore we assume $\Re(\lambda_j)\in[-\pi/2,\pi/2]$.

The state \eqref{BA2} is
an eigenstate if the Bethe equations hold:
\begin{equation}
  \label{BAe}
\left(  \frac{\sin(\lambda_j-i\eta/2)}{\sin(\lambda_j+i\eta/2)}\right)^{L}
\prod_{k\ne j}
\frac{\sin(\lambda_j-\lambda_k+i\eta)}{\sin(\lambda_j-\lambda_k-i\eta)}=1.
\end{equation}
In this case the energy is given by
\begin{equation}
\label{BAee}
  E=\sum_j e(\lambda_j),\quad\text{where}\qquad
e(u)=\frac{4\sinh^2\eta}{\cos(2u)-\cosh\eta}.
\end{equation}
In this parametrization in the regime $\Delta>1$ the one-string solutions to the Bethe
equations lie on the real axis. The Bethe equation also allows for the so-called
$n$-string solutions that are centered on the real axis:
\begin{equation}
  \lambda_k= u+i\eta(n+1-2k)/2+\delta_k\qquad k=1,\dots,n,
\end{equation}
where $u$ is the string center and $\delta_k$ are the string
deviations that become exponentially small in the thermodynamic limit.

In the thermodynamic limit the solution to the Bethe equations can be
characterized by root and hole densities $\rho_n(\lambda)$,
$\rho_n^{(h)}(\lambda)$ for the $n$-strings. It follows from the Bethe
equations 
that these functions satisfy \cite{Takahashi-book}
\begin{equation}
\label{rhoelso}
\rho_{n}+\rho^{(h)}_n= \delta_{k,1}s+
s\star \left(
  \rho^{(h)}_{n-1}+
\rho^{(h)}_{n+1}
\right),
\end{equation}
where
\begin{equation}
\label{sdef}
s(u)=1+2\sum_{n=1}^\infty \frac{\cos(2n u)}{\cosh(\eta n)},
\end{equation}
and the convolution of two functions is defined as
\begin{equation}
  (f\star g)(u)=
\int_{-\pi/2}^{\pi/2} \frac{d\omega}{2\pi} f(u-\omega) g(\omega).
\end{equation}

\subsection{Integrable initial states}
 
Given an initial state $\ket{\Psi_0}$ we are interested in the normalized squared overlaps 
 \begin{equation}
  \frac{|\skalarszorzat{\{\lambda\}_N}{\Psi_0}|^2}
   {{\skalarszorzat{\{\lambda\}_N}{\{\lambda\}_N}}}.
\end{equation}

Before starting any calculations one has to decide which initial states to consider. The
choice of $\ket{\Psi_0}$ can be motivated by their experimental relevance,
their relation to other problems in mathematical physics
\cite{zarembo-neel-cite1,zarembo-neel-cite2,zarembo-neel-cite3}, or 
their inherent integrability
properties. In the present work we follow the latter approach and
consider a special class of states that were called ``integrable
initial states'' in the recent work
\cite{sajat-integrable-quenches}. They include the N\'eel and dimer
states, and other finitely entangled Matrix Product States
(MPS's). These states are relevant both for an experimental realization and
for the AdS/CFT conjecture. Nevertheless we concentrate on their
integrability properties. 

Integrable initial states are defined as states that are
annihilated by all odd (with respect to space reflection) conserved
charges of the model \cite{sajat-integrable-quenches}:
\begin{equation}
  \label{integrablepsi}
  Q_{2j+1} \ket{\Psi_0}=0.
\end{equation}
Here the charges are defined in the usual way from the transfer matrix
of the model \cite{sajat-integrable-quenches}, or alternatively, they
can be generated in a formal way using the so-called boost operator
\cite{GM-higher-conserved-XXZ}. The eigenvalues of the charges on
Bethe states are additive:
\begin{equation}
  Q_{2j+1}\ket{\{\lambda_N\}}=\sum_{k=1}^N q_{2j+1}(\lambda_k),
\end{equation}
where $q_{2j+1}(\lambda)$ are known functions satisfying
$q_{2j+1}(\lambda)=-q_{2j+1}(-\lambda)$. 
It follows from \eqref{integrablepsi}
that the only non-vanishing overlaps are those where the Bethe
rapidities display the pair structure:
\begin{equation}
  \label{pair1}
  \{\lambda\}_N=\{\pm\lambda^+\}\cup \{\lambda^{\mathcal{S}}\},
\end{equation}
where $\lambda^{\mathcal{S}}$ are special rapidities for which
$q_{2j+1}(\lambda^{\mathcal{S}})=0$. For $\Delta>1$ we have $\{\lambda^{\mathcal{S}}\}\subset\{0,\pi/2\}$.

It was shown in \cite{sajat-integrable-quenches} that a subclass of
integrable states are those two-site states that are generated by local
$K$-matrices of the boundary Algebraic Bethe Ansatz.  The
construction of \cite{sajat-integrable-quenches} can be carried out
for an arbitrary integrable spin chain with local dimension $d\ge 2$,
if the $R$-matrix admits an appropriate crossing symmetry transformation.
Integrable states are given as
\begin{equation}
\ket{\Psi_0}=\otimes_{j=1}^{L/2} \ket{\psi},
\end{equation}
where
\begin{equation}
  \label{localtwosite}
  \ket{\psi}=\sum_{j_1,j_2=1}^d (K(-\eta/2)V)_{j_1,j_2}\ket{j_1}\otimes\ket{j_2},
\end{equation}
where $K(u)$ is a solution to the reflection equations \cite{Cherednik:1985vs,sklyanin-boundary}
\begin{equation}
  R_{1,2}(u-w)K_{1}(u)R_{1,2}(u+w)K_{2}(w)=
K_{2}(w)R_{1,2}(u+w)K_{1}(u)R_{1,2}(u-w)\,,
\label{eq:reflection_equation}
\end{equation}
and the matrix $V$ is a similarity transformation describing crossing
of the fundamental $R$-matrix:
\begin{equation}
  R^{t_1}_{1,2}(u)=\gamma(u) V^{-1}_1R_{1,2}(-u-\eta)V_1\,,
\label{eq:crossin_relation}
\end{equation}
with $\gamma(u)$ being a known function, which depends on the overall
normalization of the $R$-matrix. In the spin-1/2 chain we have $V=\sigma^y$ and the solutions to
\eqref{eq:reflection_equation} form a 3-parameter family of matrices  \cite{general-K-XXZ}.
We use the parametrization
\begin{equation}
  \label{Kparam}
\begin{split}
  K_{11}(u,\alpha,\beta,\theta)=& 2(\sinh(\alpha)\cosh(\beta)\cosh(u)+ \cosh(\alpha)\sinh(\beta)\sinh(u))\\
 K_{12}(u,\alpha,\beta,\theta)=& e^\theta \sinh(2u) \\
 K_{21}(u,\alpha,\beta,\theta)=& e^{-\theta}\sinh(2u) \\
 K_{22}(u,\alpha,\beta,\theta)=& 2(\sinh(\alpha)\cosh(\beta)\cosh(u)- \cosh(\alpha)\sinh(\beta)\sinh(u)).
\end{split}
\end{equation}
It follows that every two-site product state is integrable, because
for each $\ket{\psi}$ there are appropriate $(\alpha,\beta,\theta)$
parameters reproducing it (apart from the irrelevant overall phase and
normalization)\footnote{It was shown in \cite{sajat-integrable-quenches}
  that for other models (for example the spin-1 XXZ model) the local
  $K$-matrices produce only a subclass of two-site states, therefore
 generally not all two-site product states are integrable.}. For the
sake of completeness we give here the explicit formula for the
two-site block:
\begin{equation}
  \label{psyparam}
\begin{split}
 \psi_{11}(\alpha,\beta,\theta)=& -e^\theta \sinh(\eta) \\
  \psi_{12}(\alpha,\beta,\theta)=& 2(-\sinh(\alpha)\cosh(\beta)\cosh(\eta/2)+ \cosh(\alpha)\sinh(\beta)\sinh(\eta/2))\\
 \psi_{21}(\alpha,\beta,\theta)=& 2(\sinh(\alpha)\cosh(\beta)\cosh(\eta/2)+ \cosh(\alpha)\sinh(\beta)\sinh(\eta/2))\\
 \psi_{22}(\alpha,\beta,\theta)=&e^{-\theta}\sinh(\eta),
\end{split}
\end{equation}
where we neglected an irrelevant factor of $(i)$.

In \cite{sajat-integrable-quenches} it was also shown that a wider
class of integrable initial states can be generated by acting with the
fundamental or fused transfer matrices on the two-site states. These
states can be represented as Matrix Product States (MPS) with finite
bond dimension, therefore they have finite entanglement and can
approximate ground states of gapped Hamiltonians. However, these
states will not be considered here, we will only focus on the local
two-site states.

\subsection{Previous exact results for the finite volume overlaps}

\label{sec:prev}

Previous results in the literature concern the N\'eel and Dimer
states, where the two-site blocks are given by
\begin{equation}
  \ket{\psi_{\text{N\'eel}}}=\ket{\uparrow\downarrow}\qquad
   \ket{\psi_{\text{Dimer}}}=\frac{1}{\sqrt{2}} \left(\ket{\uparrow\downarrow}-\ket{\downarrow\uparrow}\right).
\end{equation}
They belong to the class of generalized dimer states
\begin{equation}
  \label{gendim}
  \ket{\psi_\gamma}\sim
  \left(\ket{\uparrow\downarrow}-\gamma\ket{\downarrow\uparrow}\right),\qquad
    \gamma\in \complex.
\end{equation}
These states are generated by the diagonal $K$-matrices, which can be
obtained from \eqref{Kparam} through the $\beta\to\infty$
limit. Alternatively, they can be described by the parametrization
\begin{equation}
  \label{Kdiag}
  K(u,\xi)=
  \begin{pmatrix}
    \sinh(\xi+u) & 0 \\ 0 & \sinh(\xi-u)
  \end{pmatrix}.
\end{equation}
It follows from \eqref{localtwosite} that
$\gamma=\frac{\sinh(\xi+\eta/2)}{\sinh(\xi-\eta/2)}$; the N\'eel and
Dimer states are obtained by setting $\xi=-\eta/2$ and $\xi=i\pi/2$, respectively.

A special property of these states is that every two-site block has exactly one
down spin, and so for non-vanishing overlaps the number of particles
is always $N=L/2$. Moreover, the structure of the overlaps is
essentially the same for arbitrary $\xi$: a simple argument \cite{sajat-neel,sajat-QA-GGE-hosszu-cikk} based on
the coordinate Bethe Ansatz wave function shows that for generic $\gamma$
\begin{equation}
  \label{Kidentity}
  \skalarszorzat{\Psi_\gamma}{\{\lambda\}_N}=  \skalarszorzat{\Psi_{\text{N\'eel}}}{\{\lambda\}_N}
\times \prod_{j=1}^N \frac{1+\gamma\frac{\sin(\lambda_j-i\eta/2)}{\sin(\lambda_j+i\eta/2)}}{\sqrt{1+|\gamma|}^2}.
\end{equation}
Therefore, it is enough to determine the overlaps with the N\'eel
state. The first results for this problem appeared in
\cite{sajat-neel}, where the following off-shell formula was derived
for the un-normalized overlap:
\begin{equation}
\label{N1}
  \skalarszorzat{\Psi_{\text{N\'eel}}}{\lambda_1,\dots,\lambda_N}=
\frac{\prod_j \sinh^{L}(\lambda_j-\eta/2)\sinh^{L+1}(\lambda_j+\eta/2)}
{\prod_j \sinh(2\lambda_j) \prod_{j<k} \sinh(\lambda_j-\lambda_k) \sinh(\lambda_j+\lambda_k)}
\times \det L,
\end{equation}
where $L$ is a $N\times N$ matrix with elements given by
\begin{equation}
\label{benares4}
\begin{split}
  L_{jk}=q_{2j}(\lambda_k),\quad\text{where}\qquad
q_a(u)=\coth^a(u-\eta/2)-\coth^a(u+\eta/2).
\end{split}
\end{equation}  
It is important that \eqref{N1} does not make use of the Bethe
equations,
and it merely represents an algebraic reorganization of the expression
that can be obtained directly from the wave function \eqref{BA2}. The
idea behind the derivation of \eqref{N1} is that the overlap can
be related to a specific partition function of the six-vertex model,
where the initial state $\ket{\Psi_0}$ plays the role of a reflecting
boundary \cite{sajat-neel}. For diagonal $K$-matrices this partition
function was expressed by Tsushiya as a determinant
\cite{tsushiya}, and \eqref{N1} is a specific homogeneous limit of the
Tsushiya determinant.

Despite its compact and explicit form, \eqref{N1} is not convenient
for practical purposes. First of all it is not clear how to perform the
thermodynamic limit of the determinant, and second,
the expression \eqref{N1} becomes singular for the physically relevant
cases \eqref{pair1}.

A much more useful representation for on-shell states was obtained in
\cite{Caux-Neel-overlap1,Caux-Neel-overlap2} based on an intermediate
result for the Tsushiya determinant given in \cite{sajat-karol}. It was first shown in the work
\cite{Caux-Neel-overlap1} that the overlap is non-vanishing only for
states with the pair structure, and for these cases it was found that
\begin{equation}
  \frac{|\langle \Psi_{\text{N\'eel}}|\{\pm\lambda^+\}_{N/2}\rangle|^2}
  {\langle\{\pm\lambda^+\}_{N/2}|\{\pm\lambda^+\}_{N/2}\rangle}
  =\prod_{j=1}^{N/2}  v_{\text{N\'eel}}(\lambda_j)
  \times \frac{\det_{N/2}G_{jk}^{+}}{\det_{N/2}G_{jk}^{-}}\,,
  \label{N2}
\end{equation}
where
\begin{equation}
  \label{vNeel}
  v_{\text{N\'eel}}(\lambda)=\frac{{\tan(\lambda_{j}+i\eta/2)\tan(\lambda_{j}-i\eta/2)}}
    {4\sin^2(2\lambda_{j})}
\end{equation}
and\footnote{The matrix elements of $G^{\pm}$
include a sign difference as compared to the formulas given in
\cite{Caux-Neel-overlap1,Caux-Neel-overlap2}, but they agree with
those of \cite{korepin-norms,Korepin-Izergin-XXZ}. This sign does not
influence the resulting overlaps. We introduced it in order to
have positive elements in the diagonal of the matrix, and to avoid
confusion with earlier results when taking the thermodynamic limit.
}
\begin{equation}
  \label{gaudins}
  G_{jk}^{\pm}  =  \delta_{jk}\left(-L\varphi_{\eta/2}(\lambda^+_{j})
    +\sum_{l=1}^{L/4}\varphi_{\eta}^{+}(\lambda^+_{j},\lambda^+_{l})\right)
  -\varphi^{\pm}_\eta(\lambda^+_{j},\lambda^+_{k})
  \end{equation}
  with
  \begin{equation}
    \label{varphix}
    \begin{split}
\varphi^{\pm}_\eta(\lambda,\mu) & =  \varphi_\eta(\lambda-\mu)\pm
\varphi_\eta(\lambda+\mu)\\
\varphi_x(\lambda) & = \frac{\sinh(2x)}{\sinh(\lambda+ix)\sinh(\lambda-ix)}\,.
\end{split}
\end{equation}
In \eqref{N2} it was assumed that $N$ is even and the Bethe state does
not include the special rapidities
$\{\lambda^{\mathcal{S}}\}\subset\{0,\pi/2\}$; for those cases
different regularized formulas are needed
\cite{Caux-Neel-overlap2,finite-qa}. 

Based on the identity \eqref{Kidentity} it is possible to derive the
overlaps with the other states generated by diagonal $K$-matrices, and
they all take the form of \eqref{N2} with a
different single particle overlap function.
For example for the Dimer state we get
\cite{sajat-oTBA,sajat-QA-GGE-hosszu-cikk}
\begin{equation}
  v_{\text{Dimer}}(\lambda)=\frac{\sinh^4(\eta/2)\cot^2(\lambda)}
  {\sin(2\lambda+i\eta)\sin(2\lambda-i\eta)}.
\end{equation}

Further overlap formulas were computed in the works
\cite{zarembo-neel-cite1,zarembo-neel-cite2,zarembo-neel-cite3}. These
papers considered certain Matrix Product States (MPS's) in the $SU(N)$
symmetric models, that are relevant for the AdS/CFT conjecture.
Regarding the case of the spin-1/2 XXX chain
real space calculations were used to derive the overlaps with Bethe states
with a low number of particles, and formulas of the form \eqref{N2}
were conjectured for higher particle numbers.
It was shown in \cite{sajat-integrable-quenches} that the MPS's in
question are in fact integrable, and they can be identified
either as zero-momentum components of specific two-site states, or
they result as action of transfer matrices on such states.
Exact overlap formulas were derived in \cite{zarembo-neel-cite2},
which considered overlaps with the so-called partial N\'eel states. They
are zero-momentum components of two-site states where
$\psi_{11}=0$. According to \eqref{psyparam} they correspond to a
special scaling limit, where the $K$-matrix becomes lower
diagonal. For these cases an exact off-shell formula was derived, that
is closely related to \eqref{N1}, with a modified matrix. However, in
the on-shell case an exact result of the form \eqref{N2} was
obtained once again. The work \cite{zarembo-neel-cite2} used a
slightly modified version of the argument of Tsushiya \cite{tsushiya}
to obtain the off-shell result, but it was demonstrated in
\cite{calabrese-recursive-overlaps} that it is not evident how to
apply this idea to the generic case with $\psi_{11}\ne 0$.

The results of 
\cite{zarembo-neel-cite1,zarembo-neel-cite2,zarembo-neel-cite3} seem
to indicate that the on-shell overlaps are always of the form
\eqref{N2}, even for off-diagonal $K$-matrices. However, this is
somewhat misleading. These works only consider the $SU(2)$ invariant
chain, where every $K$ matrix can be rotated to diagonal one. 
It is known that the Bethe vectors are highest weight
states and that the spin lowering operators can be represented with
infinite rapidity particles. Therefore, the $SU(2)$ rotations can be
evaluated on the Bethe vectors, and overlaps with off-diagonal
$K$-matrices can be expressed with those of the known diagonal
cases. This procedure is analogous to that used in
\cite{Brockmann-BEC} to study $q$-raised N\'eel states in the XXZ
model. Nevertheless we believe that the method of
\cite{zarembo-neel-cite2} could be applied to lower diagonal
$K$-matrices even in the XXZ case, which would then yield genuine
new rigorous results.

Further results appeared in \cite{ADSMPS2}, which considered MPS's for
higher rank models. In the  $SU(3)$ case formulas
analogous to \eqref{N2} were obtained from coordinate Bethe Ansatz, with the determinants mirroring the nested
Bethe Ansatz solution of the model. The resulting overlap formulas
were used in \cite{nested-quench-1} to study the corresponding
quench situation, and it was later shown in
\cite{sajat-integrable-quenches} that the initial state is in fact
integrable. On the other hand, the interpretation of these results
within Boundary Algebraic Bethe Ansatz is not yet known, and this is
an intriguing problem to be investigated in future work. However, in
the present paper we content ourselves with the open questions of the spin-1/2 XXZ chain. 

\subsection{The Quench Action method}

\label{sec:QA}

The Quench Action (QA) method \cite{quench-action} was devised to
determine which Bethe states are relevant for the long-time behavior of
physical observables in quantum quenches. The essence is to
investigate the overlap sum rule
\begin{equation}
  \label{ovsumrule}
  1=\skalarszorzat{\Psi_0}{\Psi_0}=
  \sum_{\{\lambda\}}
 \frac{|\langle \Psi_0|\{\lambda\}\rangle|^2}
  {\langle\{\lambda\}|\{\lambda\}\rangle},
\end{equation}
where the summation runs over all eigenstates of a finite volume
system, and to select the states that dominate the sum in the
thermodynamic limit. It can be argued that the same states determine
the long-time averages of local observables \cite{quench-action}.
The selection of the relevant states is achieved by transforming the
finite sum into a functional integral over root densities,
and minimizing the resulting Quench Action, which is defined as the combination 
of the overlap and entropy of a state with a given root distribution. 
The QA is completely analogous to the free energy
functional of the finite temperature situation, therefore the resulting
equations always take the form of (generalized) Thermodynamic
Bethe Ansatz (TBA) equations \cite{Takahashi-book}.

Before exchanging the summation for a functional integral it is
important to remember that for the integrable quenches considered here
the non-vanishing terms in the sum \eqref{ovsumrule}
are the parity invariant Bethe states. 
Therefore, these states can be completely determined by listing the
positive rapidities (defined either as $\Re(\lambda^+)>0$ for all roots, or
$\lambda_n^+>0$ for the string centers), and this affects the entropy
associated to a given root configuration. For the $\ordo(L)$ term of
the entropy in the QA we get
 a factor of $1/2$ as compared to the formulas of the usual TBA.
In the next subsection we also investigate the $\ordo(1)$ pieces,
but first we collect the known formulas for the leading part. 

In \cite{quench-action} it was shown that the sum \eqref{ovsumrule}
leads to the functional integral
\begin{equation}
  \int D\rho_n(\lambda^+) e^{-S_{QA}(\{\rho_n(\lambda)\})}
\end{equation}
with
\begin{equation}
  \label{QAdef}
  S_{QA}(\{\rho_n(\lambda)\})=
-2\Re\ln\langle\Psi_0| \{\rho_n(\lambda)\}  \rangle +\frac{1}{2}S_{YY}(\{\rho_n(\lambda)\})
\end{equation}
with $S_{YY}$ being the Yang-Yang entropy
\begin{equation}
 S_{YY}({\rho_n(\lambda)})\equiv L\sum_{n=1}^\infty\int_{-\infty}^{+\infty}d
\lambda\Big[\rho_n(\lambda)\ln\Big(1+\frac{\rho_n^h(\lambda)}{\rho_n(\lambda)}
\Big)+\rho_n^h(\lambda)\ln\Big(1+\frac{\rho_n(\lambda)}{\rho^h_n(\lambda)}\Big)
\Big].
\end{equation}
The factor of $1/2$ in front of $S_{YY}$ results from the aforementioned restriction on
parity invariant states. 

In writing down \eqref{QAdef} we assumed that the overlap is a known smooth function of the
root densities. In particular, the TBA formalism can be applied if
it can be written as
\begin{equation}
  \label{ext}
  \lim_{L\to\infty} \frac{1}{L} 2\Re\ln\langle\Psi_0| \{\rho_n(\lambda)\}\rangle
  =\sum_{n=1}^\infty \int d\lambda \rho_n(\lambda) g_n(\lambda),
\end{equation}
with $g_n(\lambda)$ being the $n$-string overlap functions. If
\eqref{ext} holds for most Bethe states in the thermodynamic limit (TDL), then the
minimization of the Quench Action leads to the following set of
generalized TBA equations \cite{JS-oTBA,enej-gge}:
\begin{equation}
\label{oTBA}
\log Y_j=d_j +s\star \left[
\log(1+Y_{j-1})+\log(1+Y_{j+1})
\right],
\end{equation}
where $Y_j(\lambda)=\rho^{(h)}_j(\lambda)/\rho_j(\lambda)$ and
\begin{equation}
  \label{dj}
d_j =-g_j+s\star(g_{j-1}+g_{j+1}), \text{ with } g_0=0.
\end{equation}
Supplied with asymptotic conditions on the growth of $Y_j$ as
$j\to\infty$ these equations completely determine the root densities. 

We note that for a generic initial state there would be no guarantee that the overlaps lead to
the form \eqref{ext} in the TDL, and even if they do, the $g_n(\lambda)$
functions could be algebraically independent from each other.
In fact, it is known that an analogous situation happens for
non-integrable states in the
framework of GGE \cite{enej-gge}. On the other hand, a relatively
simple situation arises if the overlap factorizes 
in finite volume.

We call an overlap factorizable if for on-shell states it can be written as
\begin{equation}
  \label{factorized}
  \frac{|\langle \Psi_{0}|\{\pm\lambda^+\}_{N/2}\rangle|^2}
  {\langle\{\pm\lambda^+\}_{N/2}|\{\pm\lambda^+\}_{N/2}\rangle}
=A^{L}  \prod_{j=1}^{N/2} v(\lambda_j^+)
  \times C(L,N,\{\lambda\}),
\end{equation}
where $C(L,N,\{\lambda\})$ is a function that remains $\ordo(L^0)$ in
the thermodynamic limit. 
The known exact results presented in subsection \ref{sec:prev} were
all factorizable.
The separate pre-factor $A^L$ was not present in previous results,
but it can be argued that such a factor necessarily appears 
for initial states having components with non-zero magnetization,
in which cases the particle number is independent of $L$. In
\eqref{factorized} we assumed the pair structure;
 we will argue below that factorizability is a property that
indeed only holds for integrable initial states. 

It is important that the formula \eqref{factorized} represents
factorization on the level of the individual rapidities. For the
string contributions in \eqref{ext} we obtain
\begin{equation}
  \label{gj}
 g_n(\lambda)=-\sum_{k=1}^n
  \log\big(v(\lambda+i\eta(n+1-2k)/2)\big),
\end{equation}
and this specifies the sources in \eqref{oTBA} through  \eqref{dj}.
As a consequence, it can be shown that the solutions of the QA-TBA
always satisfy the $Y$-system equations
\begin{equation}
  \label{Y}
  Y_j(\lambda+i\eta/2) Y_j(\lambda-i\eta/2)=(1+ Y_{j-1}(\lambda))(1+ Y_{j+1}(\lambda)),
\end{equation}
The proof is straightforward by combining \eqref{oTBA}, \eqref{dj} and
\eqref{gj} and making use of the relation
\begin{equation}
  \label{identity}
\lim_{\delta\to\eta/2}\left[  \int_{-\pi/2}^{\pi/2}\frac{dx}{2\pi}
  (s(y+i\delta-x)+s(y-i\delta-x)) f(x)\right]= f(y),
\end{equation}
which follows from the integral representation \eqref{sdef}
and is valid for any smooth function $f(x)$.

The Y-system is
regarded as an important sign of integrability of the initial state,
and it was used as a tool for finding exact solution to the TBA
\cite{JS-oTBA,jacopo-massless-1}. In the
Quench Action framework the Y-system ultimately follows from factorizability of
the overlaps, which is expected to be a general property of integrable
initial states \cite{sajat-integrable-quenches}.
Conversely, in the generic case where
the $Y$-system does not hold \cite{JS-CGGE,enej-gge}, the overlaps
with the initial state can not have the factorized form.

\subsection{The overlap sum rule}

\label{sec:ovsumrule}

It is worthwhile to investigate the overlap sum rule \eqref{ovsumrule} in more
detail, which sheds some light on the role of the Gaudin-like determinants in
the overlaps. To this order we give a slightly different definition of the
Quench Action.

In analogy with the partition functions in thermodynamics we define
\begin{equation}
  \label{QAdef2}
  S_{QA}=-\log \sum_{\{\pm\lambda^+\}}
   \frac{|\langle \Psi_0|\{\pm\lambda^+\}_{N/2}\rangle|^2}
  {\langle\{\pm\lambda^+\}_{N/2}|\{\pm\lambda^+\}_{N/2}\rangle},
\end{equation}
with a large volume behavior given by
\begin{equation}
  S_{QA}=s_{QA}L+\Delta S+\ordo(L^{-1}),
\end{equation}
where $s_{QA}$ is the QA density, and $\Delta S$ is an $\ordo(1)$
piece that has not yet been investigated in the literature. 
From the definition we have the trivial identity $S_{QA}=0$, and our goal is to
derive this within the Quench Action method, both for the leading and
the sub-leading part.

The extensive part is given by \eqref{QAdef}-\eqref{ext} evaluated at
the saddle point solution. It was derived in
\cite{sajat-oTBA,sajat-QA-GGE-hosszu-cikk} that
\begin{equation}
  \label{ovsumext}
  s_{QA}= \int_{-\pi/2}^{\pi/2}
  \frac{d\lambda}{2\pi} s(\lambda)
  \left[ \log(1+Y_1(\lambda)) +\log(v(\lambda))\right],
\end{equation}
where $Y_j(\lambda)$ is the solution of the QA-TBA \eqref{oTBA} and
$v(\lambda)$ is the single-particle overlap function.
Here it was assumed that the initial state has zero total magnetization\footnote{The generic case was not
discussed in \cite{sajat-oTBA,sajat-QA-GGE-hosszu-cikk}, because existing overlap formulas were only
available for initial states with zero total magnetization.}; the
formula is
analogous to the expression of the free energy density within the
usual TBA \cite{Takahashi-book}. In
\cite{sajat-oTBA,sajat-QA-GGE-hosszu-cikk} the relation $s_{QA}=0$ was
used as a consistency condition within the QA; it also served as a
 test for the accuracy of the numerical solution of the TBA equations.

In order to determine the finite term $\Delta S$ we first investigate the thermodynamic limit of the
overlap formula \eqref{N2}. Previously we have treated the extensive
part, which leads to \eqref{ext} with the $n$-string overlaps given by
\eqref{gj}. On the other hand, the ratio of the determinants in
\eqref{N2} gives a non-zero $\ordo(1)$ contribution.
In order to evaluate this term the first step is to express the
ratio of determinants using only the string centers. This task is
analogous to finding the norm of Bethe states with strings \cite{kirillov-korepin-norms-strings}, and for
the determinants in question it was performed in
\cite{finite-qa}. Denoting by $\lambda_{n,a}$ the $n$-string
centers with index $a$ we get
\begin{equation}
  \frac{\det G^+}{\det G^-}\quad \to\quad  \frac{\det \tilde G^+}{\det \tilde G^-}
\end{equation}
with
\begin{equation}
  \tilde G_{(n,a),(m,b)}^{\pm} =  \delta_{(n,a),(m,b)}
  \left(-L\varphi_{n\eta/2}(\lambda_{j})
    +\sum_{(o,c)}\Theta_{n,o}^{+}(\lambda_{n,a},\lambda_{o,c})\right)-
  \Theta_{n,m}^{\pm}(\lambda_{n,a},\lambda_{m,b}),
\end{equation}
where
\begin{equation}
 \Theta_{n,m}^{\pm}(\lambda,\mu) =   \Theta_{n,m}(\lambda-\mu)\pm  \Theta_{n,m}(\lambda+\mu)
\end{equation}
\begin{equation}
  \Theta_{n,m}(\lambda)=
  \begin{cases}
    \varphi_{|n-m|\eta/2}(\lambda)
+\sum_{j=1}^{(n+m-|n-m|-1)/2} \varphi_{(|n-m|+2j)\eta/2}(\lambda)+
\varphi_{|n+m|\eta/2}(\lambda)
    & \text{ if } n\ne m\\
\sum_{j=1}^{n-1} 2\varphi_{j\eta}(\lambda)+\varphi_{n\eta}(\lambda)       & \text{ if } n=m\\
  \end{cases},
\end{equation}
and $\varphi_x(\lambda)$ is given in \eqref{varphix}.

In the thermodynamic limit the ratio of the two determinants leads to
a ratio of two Fredholm determinants. The calculation proceeds through
standard steps \cite{korepin-LL1,Korepin-Izergin-XXZ,norms-destri}
and here we only give
the main result. In the TDL we get
\begin{equation}
  \lim_{L\to\infty}   \frac{\det \tilde G^+}{\det \tilde G^-}=
  \frac{\det \left(1-\hat Q^+\right)}{\det \left(1-\hat Q^-\right)},
\end{equation}
where $\hat Q^\pm$ are integral operators that act on functions
$f_n(\lambda)$ with a string index $n$ defined for $\lambda\in
\valos^+$. The action of the integral operators reads
\begin{equation}
\label{Q+-}
  \big(\hat Q^\pm
  (f)\big)_n(x)=\sum_{m=1}^\infty \int_{0}^\infty\frac{dy}{2\pi}
\big(\Theta_{n,m}^\pm(x,y)\big)   \frac{1}{1+Y_m(y)}  f_m(y).
\end{equation}

The second step to calculate the $\ordo(1)$ terms in the sum rule is
to carefully consider the summation over the eigenstates and the
transformation of the sum into a functional integral. It was explained
in the papers \cite{sajat-g,woynarovich-uj} (inspired by the earlier
work \cite{woynarovich}) that in such situations there are two
$\ordo(1)$ contributions: one coming from the density of states in
rapidity space (corresponding to the change of variables from the
integer momentum quantum numbers to the rapidities) and one from the
quadratic fluctuations around the saddle point solution. The two terms
depend on the nature of the Bethe equations. In the usual
TBA setting with periodic boundary conditions the two contributions just
cancel each other, whereas in a system with boundaries they combine to
a finite and well defined term. In the overlap sum
rule the only allowed states are those with the pair structure, and
this corresponds formally to the boundary case, because both the
density of states and the variations around the saddle point have to
be calculated by varying only half of the rapidities. Therefore, the
boundary results of \cite{sajat-g,woynarovich-uj} apply, which read
\begin{equation}
  \sum_{\{\lambda^+,-\lambda^+\}}
  \quad
\to\quad
 \frac{\det \left(1-\hat Q^-\right)}{\det \left(1-\hat Q^+\right)}
 \int D\rho_n(\lambda^+),
\end{equation}
where $\hat Q^\pm$ are the same Fredholm that appeared as the
limits of the Gaudin-like determinants. We note that the papers
\cite{sajat-g,woynarovich-uj} did not discuss theories with multiple
particle species (such as the strings in the XXZ chain), however the
generalization of the results given there is straightforward.

Combining the previous results we find that
\begin{equation}
  e^{-\Delta S}=  \frac{\det \left(1-\hat Q^-\right)}{\det \left(1-\hat Q^+\right)}\times
\frac{\det \left(1-\hat Q^+\right)}{\det \left(1-\hat Q^-\right)}=1,
\end{equation}
as required by definition. Thus the ratio of the two Gaudin-like
determinants ensures that the overlaps have the correct normalization
in the thermodynamic limit. We believe that this has not been noticed
in earlier works.

The present result applies to the states \eqref{gendim} generated by
the diagonal $K$-matrices, where the overlap of the form \eqref{N2} is
rigorously proven. It is remarkable, that the $\ordo(1)$ terms do not
depend on the free parameter of the initial state: for the overlap
this follows from relation \eqref{Kidentity}, whereas for the
normalization of the functional integral it is a result of the pair
structure. This leads to the conjecture, that the same Gaudin-like
determinants should appear also for the generic $K$-matrices; this
conjecture and its tests are presented in Section \ref{sec:ov}. 
In the Conclusions we discuss further implications of these observations.

\subsection{Calculation of the Loschmidt echo}

An important property of integrable two-site states is that
 the extensive part of the Loschmidt amplitude (also called the
dynamical free energy) can be computed
analytically. 
The Loschmidt amplitude is defined as the overlap of a time evolved
state with the original initial state:
\begin{equation}
  \label{Loschm}
  L(t)=\bra{\Psi_0}e^{-iHt}\ket{\Psi_0},
\end{equation}
and the dynamical free energy is defined as
\begin{equation}
  g(w)=-\lim_{L\to\infty } \frac{1}{L} \log L(-iw).
\end{equation}
It was shown in \cite{sajat-BQTM,sajat-Loschm} that these quantities
 can be evaluated by a lattice path integral, where the
initial and final states $\bra{\Psi_0}$ and $\ket{\Psi_0}$ play the
role of boundary conditions.
The corresponding partition
function can be evaluated in the rotated channel by the so-called
Boundary Quantum Transfer Matrix (QTM), if the two site block
$\ket{\psi}$ is identified with two different $K$-matrices
$K_{\pm}(u,\alpha_\pm,\beta_\pm,\theta_\pm)$
as
\begin{equation}
  \label{localtwosite2}
  \ket{\psi}\sim \sum_{j_1,j_2=1}^2 (K_-(-\eta/2)\sigma^y)_{j_1,j_2}\ket{j_1}\otimes\ket{j_2},\qquad
 (\ket{\psi})^*\sim \sum_{j_1,j_2=1}^2 (\tilde K_+(\eta/2)\sigma^y)_{j_1,j_2}\ket{j_1}\otimes\ket{j_2},
\end{equation}
where $\tilde K_+$ denotes transposition.
It can be seen from \eqref{Kparam} that the two vectors in \eqref{localtwosite2} can be proportional to
each other only if the two sets of parameters satisfy
\begin{equation}
  \alpha_-=-\alpha_+^*\equiv \alpha\qquad \beta_-=\beta_+^*\equiv \beta
  \qquad \theta_-=-\theta_+^*\equiv \theta.
\end{equation}

In the thermodynamic limit the amplitude \eqref{Loschm} will be given
by the leading eigenavalue of the Boundary QTM, for which 
the following generalized TBA equations were derived in \cite{sajat-Loschm}\footnote{The
  formulas here have an additional factor of 4 in front of the energy
  terms; which originates from a different definition of the Hamiltonian.}  
\begin{equation}
\label{oTBA2}
\log \tilde y_j=- 4\sinh(\eta)w\  s\delta_{j,1}+
\tilde d_j +s\star \left[
\log(1+\tilde y_{j-1})+\log(1+\tilde y_{j+1})
\right],
\end{equation}
such that the dynamical free energy is given by
\begin{equation}
  g(w)=\frac{1}{2}\int_{-\pi/2}^{+\pi/2}
  \,d\lambda\ s(\lambda)\left\{ w 2\sinh(\eta) a(\lambda)+ \log\left[\frac{1+\tilde{y}_1(\mu)}{
1+\tilde Y_1(\lambda) }\right]\right\}\,,
\label{final_g}  
\end{equation}
where
\begin{equation}
a(\lambda)=\frac{4\sinh(\eta)}{\cosh(\eta)-\cos(2\lambda)}\,.
\label{a_function}
\end{equation}
and $\tilde Y_j(\lambda)$ is the solution of \eqref{oTBA2} for
$w=0$.
The source terms $\tilde d_j$ are independent of $w$ and are
determined by the parameters of the
$K$-matrices. A number of specific cases were discussed in
\cite{sajat-Loschm}, but we refrain from repeating the explicit
formulas. Equation \eqref{oTBA2} will be called ``Loschmit-TBA'' in
the rest of the paper.

An important central result of \cite{sajat-Loschm} was that the
functions $\tilde Y_j$
satisfy the $Y$-system equations, which follows from the fusion
hierarchy of the Boundary QTM. Also, it was shown that 
the first member is given
explicitly by\footnote{The most general case with arbitrary complex
 $(\alpha,\beta,\theta)$ parameters was not given in \cite{sajat-Loschm}. However, it is
  straightforward to extract it from the intermediate results given there.}
\begin{equation}
  \label{Y1}
  1+Y_1(\lambda)=\frac{\mathcal{N}(\lambda+i\eta/2)\mathcal{N}(\lambda-i\eta/2)}
  {\chi(\lambda)}
\end{equation}
where
\begin{equation}
  \mathcal{N}(\lambda)=\text{Tr}\Big[ K_+(\lambda+\eta/2)K_-(\lambda-\eta/2)\Big]
\end{equation}
and 
\begin{equation}
  \label{chidef}
  \chi=16 \frac{v^s_\eta v^c_\eta}{v^s_{\eta/2}v^c_{\eta/2}}
 v^s_\alpha v^s_{\alpha^*}v^c_\beta v^c_{\beta^*}
\end{equation}
where we introduced the short-hand notation
\begin{equation}
  v^s_\kappa(\lambda)=\sin(\lambda+i\kappa)\sin(\lambda-i\kappa)\qquad
   v^c_\kappa(\lambda)=\cos(\lambda+i\kappa)\cos(\lambda-i\kappa)
 \end{equation}
Note that according to \eqref{localtwosite2} $\mathcal{N}(0)$ describes the norm
of the two-site block $\ket{\psi}$. 

Based on the similarities of the QA formulas \eqref{oTBA} and those obtained for
the Loschmidt echo it is very natural to identify
\begin{equation}
  \label{connection}
  \tilde Y_j(\lambda)=Y_j(\lambda)\qquad \tilde d_j(\lambda)=d_j(\lambda),
\end{equation}
which was known to hold for the diagonal $K$ matrices \cite{sajat-QA-GGE-hosszu-cikk}
and was also checked explicitly for specific off-diagonal cases in
\cite{sajat-Loschm}. 
Eq. \eqref{connection} represents a close connection between the QA and QTM
approaches: it implies that
the integrability properties of the overlaps (i.e. factorizability)
are closely related to the fusion hierarchy of the Boundary QTM.
This is the observation that allows us to extract
new finite volume overlap formulas from the Loschmidt-TBA.

\section{The general overlap formula}

\label{sec:ov}

In the previous section it was explained that \eqref{oTBA2} is a generalized TBA
equation describing the Loschmidt-echo, valid for arbitrary
$K$-matrices, such that the solution at
$w=0$ (corresponding to the Quench Action point) is exactly known
through \eqref{Y1}. 
If we assume that there exists a factorized overlap
of the form \eqref{factorized}, then \eqref{oTBA2} has to coincide with
the QA-TBA \eqref{oTBA} for that specific overlap. This  is supported by the identification \eqref{connection}.
In this case the single-particle overlap
function can be ,,reverse engineered'' by computing the overlap
sources $d_j(u)$ in \eqref{oTBA} and comparing them to $\tilde d_j(u)$
in \eqref{oTBA2}. Instead of computing the source terms, we choose to
operate only with the $Y$-functions and their singularity properties.

For example, for the first equation in the TBA
we obtain the suggestive formula
\begin{equation}
  \label{Y1v}
  \begin{split}
&  \log Y_1(\lambda)
-s\star  \Big[\log(Y_1(\lambda+i\eta/2))+\log(Y_1(\lambda-i\eta/2))\Big]
=\\ &
\hspace{2cm}=-\log(v(\lambda))+s\star
  \Big[\log(v(\lambda+i\eta/2))+\log(v(\lambda-i\eta/2))\Big].
 \end{split} 
\end{equation}
Here we used the $Y$-system equations \eqref{Y}, and relations \eqref{dj} and
\eqref{gj} for the overlaps. It is tempting to identify
$v(\lambda)=C/Y_1(\lambda)$, however, this is misleading. Relation
\eqref{Y1v} gives information only about the poles and zeroes 
of the functions within the physical
strip $|\Im{(\lambda)}|\le \eta/2$. This follows from the identity
\begin{equation}
  \label{idill}
    \int_{-\pi/2}^{\pi/2} \frac{dx}{2\pi} s(y-x)
    \log(h(x+i\eta/2)h(x-i\eta/2))=\log h(y),
\end{equation}
which can be obtained from the Fourier representation of $s(\lambda)$
\eqref{sdef}, or alternatively after a contour shift from \eqref{identity},
assuming that there are no singularities of $\log(h(x))$ to be picked up. Therefore,
\eqref{Y1v} implies that $v(\lambda)=h(\lambda)/Y_1(\lambda)$, where
$h(\lambda)$ is free of poles and zeroes in the physical strip, and 
the poles (or zeroes) of $Y_1(\lambda)$ within the physical strip
correspond to zeroes (or poles) of $v(\lambda)$, respectively. 

In order to understand the dependence of the overlaps on the
$K$-matrices it is useful to study the known cases. 
For a diagonal $K$-matrix of the form \eqref{Kdiag} the
overlap function is 
\begin{equation}
  \label{diagv}
  v(\lambda)=
  \frac{\sinh^4(\eta)}{16 (\cosh(2\xi)\cosh(\eta)-1)^2}
\frac{(v^s_\xi)^2}{v^s_0 v^c_0 v^s_{\eta/2} v^c_{\eta/2}},
\end{equation}
which is obtained from the relation \eqref{Kidentity} and the N\'eel
overlap function \eqref{vNeel}. For $Y_1(u)$
the following result holds:
\begin{equation}
  \label{xiY}
  1+Y_1(\lambda)=(1+\fa(\lambda+i\eta/2)(1+\fa^{-1}(\lambda-i\eta/2))
\end{equation}
with 
\begin{equation}
  \fa(\lambda)=
\frac{\sin(\lambda-i(\xi-\eta/2))}{\sin(\lambda+i(\xi-\eta/2))}
\frac{\sin(\lambda+i(\xi+\eta/2))}{\sin(\lambda-i(\xi+\eta/2))}
\frac{\sin(2\lambda-i\eta)}{\sin(2\lambda+i\eta)}.
\end{equation}
Eq. \eqref{xiY} could be obtained by taking the diagonal limit of the
general formula \eqref{Y1}, or from earlier results for the diagonal
$K$-matrices \cite{sajat-BQTM,sajat-QA-GGE-hosszu-cikk}. It can be
seen that the only poles of $Y_1(\lambda)$ within the physical strip can be at
$\lambda=\pm \xi$, if $\Re(\xi)<\eta/2$, and these poles are indeed
reflected by the zeroes of $v(\lambda)$. In the cases where $\xi$ is
outside the physical strip, the two singularities still move together,
which is ensured by analytic continuation. The poles of $v(\lambda)$ within
the physical strip are the double poles at $\lambda=0,\pi/2$, and direct
calculation shows that indeed they correspond to zeroes of
$Y_1(\lambda)$. There are additional poles of both $v(u)$ and $Y_1(\lambda)$ at $\lambda=\pm
i\eta/2$ and $\lambda=\pi/2 \pm i\eta/2$, but they lie on the boundary of
the physical strip, and \eqref{Y1v} does not give any information
about them: the contributions of symmetrically located singularities at $\kappa\pm
i\eta/2$ cancel each other in \eqref{Y1v} for any $\kappa\in\valos$,
but they are relevant  for the higher nodes of the TBA.
The crucial observation here is that there are some fixed
($\xi$-independent) singularities of both functions, and extra zeroes of $v(\lambda)$ that are
determined by the $\xi$-dependent poles of $Y_1(\lambda)$. This is enough to
conjecture the overlap function for  the general case.

For a generic $K$-matrix the poles of $Y_1$ are given by
\eqref{chidef}, and comparing to the previous result \eqref{diagv}
we arrive at the conjecture
\begin{equation}
  \label{ulambda}
  v(\lambda)=Cu(\lambda),\qquad u(\lambda)=\frac{  v_{\alpha}^sv_{\alpha^*}^sv_{\beta}^cv_{\beta^*}^c}
  {v^s_{\eta/2} v^c_{\eta/2}v^s_0v^c_0},
\end{equation}
where $C$ is a numerical constant that depends on the parameters
$(\alpha,\beta,\theta)$. A detailed analysis of all higher nodes in
the TBA reveals that this overlap
function indeed reproduces the correct source terms.
We note that
\eqref{ulambda} does not include any ``minimal analyticity
assumption'': any other function $\tilde v(\lambda)$ has to have
exactly the same poles and zeroes as $v(\lambda)$ in 
order to yield the correct Loschmidt-TBA, and combined with the
symmetry properties we get $\tilde v(\lambda)=\tilde C
v(\lambda)$.

The overall normalization plays a similar role as a magnetic
field in the thermodynamics: it influences the net magnetization.
For states with zero total magnetization the constant $C$ can be fixed
by the overlap sum rule. In these cases we expect that $A=1$ in
\eqref{factorized}, and the extensive part of the overlap sum rule
\eqref{ovsumext} gives
\begin{equation}
  \int_{-\pi/2}^{\pi/2} \frac{d\lambda}{2\pi} s(\lambda) \log\left[
    \mathcal{N}(\lambda+i\eta/2)\mathcal{N}(\lambda-i\eta/2)
\frac{C(\alpha,\xi)}
  {16v^s_\eta(\lambda) v^c_\eta(\lambda) v^s_0(\lambda)v^c_0(\lambda)}
  \right]=0.
\end{equation}
Substituting $h(\lambda)=\sin(2\lambda+i\eta)\sin(2\lambda-i\eta)$ into \eqref{idill} we get
\begin{equation}
  \int_{-\pi/2}^{\pi/2} \frac{d\lambda}{2\pi} s(\lambda)
  \log \frac{1}{16v^s_\eta(\lambda) v^c_\eta(\lambda) v^s_0(\lambda)v^c_0(\lambda)}
=-\log \sinh^2(\eta).
  \end{equation}
It can be checked that there are no zeros of $\mathcal{N}$ within the
physical strip (this is most easily checked for the diagonal case
\eqref{Kdiag}), which implies
\begin{equation}
  \int_{-\pi/2}^{\pi/2} \frac{d\lambda}{2\pi} s(\lambda) \log
  ( \mathcal{N}(\lambda+i\eta/2)\mathcal{N}(\lambda-i\eta/2))=
  \log \mathcal{N}(0).
\end{equation}
From this we get the result
\begin{equation}
  \label{leoszt}
  C(\alpha,\beta,\theta)=\frac{\sinh^4(\eta)}{\mathcal{N}^2(0)}.
\end{equation}

It is important that the previous arguments only fix the thermodynamic
part of the overlap. On the other hand,
Section \ref{sec:ovsumrule}  showed that the two
Gaudin-like determinants produce just the correct normalization in
order to satisfy the overlap sum rule, and the $\ordo(1)$ terms do not
depend on the extensive part, they are fixed simply by the pair
structure. 
Therefore it is tempting to assume that the finite part is
always given by the ratio of the same Gaudin-like determinants. This leads
to the general finite volume conjecture
\begin{equation}
  \frac{|\langle\Psi_{0}|\{\pm\lambda^+\}_{N/2}\rangle|^2}
  {\langle\{\pm\lambda^+\}_{N/2}|\{\pm\lambda^+\}_{N/2}\rangle}
  =\tilde A^{L-2N}\prod_{j=1}^{N/2}  v(\lambda^+_j)
  \times \frac{\det_{N/2}G_{jk}^{+}}{\det_{N/2}G_{jk}^{-}}\,,
  \label{OVERLAPS}
\end{equation}
with the matrices given by \eqref{gaudins}. As remarked in
Sec. \ref{sec:QA}, the pre-factor
$\tilde A^{L-2N}$ can be present for states with non-zero total
magnetization. If formula \eqref{OVERLAPS} is correct, then it should
also apply to the zero particle case, which fixes
$\tilde A$:
\begin{equation}
  \tilde A=
\frac{|\psi_{11}|}{|\psi|}=
  \frac{|K^-_{12}(-\eta/2)|}{\sqrt{|\mathcal{N}(0)|}}.
\end{equation}
Combined with \eqref{psyparam} this gives the remarkably simple formula
\begin{equation}
  \frac{|\langle\Psi_{0}|\{\pm\lambda^+\}_{N/2}\rangle|^2}
  {\langle\{\pm\lambda^+\}_{N/2}|\{\pm\lambda^+\}_{N/2}\rangle}
  =
  \frac{|e^{\theta(L-2N)}|\sinh^{L}(\eta)}{|\mathcal{N}(0)|^{L/2}}
\prod_{j=1}^{N/2}  u(\lambda_j)
  \times \frac{\det_{N/2}G_{jk}^{+}}{\det_{N/2}G_{jk}^{-}}\,,
  \label{OVERLAPS2}
\end{equation}
with $u(\lambda)$ given by \eqref{ulambda}.

It is useful to give a few comments about the formula \eqref{OVERLAPS2}.
The
denominator $|\mathcal{N}(0)|^{L/2}$ can be interpreted as an overall
normalization factor coming from the norm of a single two-site block
$\ket{\psi}$, as given by \eqref{localtwosite}.
The $\theta$ parameter of the $K$-matrix
does not appear in the function $u(\lambda)$, but it affects the
norm $\mathcal{N}(0)$ and it appears in a separate pre-factor. This factor can 
be easily understood from the
coordinate Bethe Ansatz: it can be seen from
\eqref{psyparam} that in the un-normalized vector
$\ket{\Psi_0}$ the up/down spin components carry factors of
$e^{\pm\theta/2}$, respectively; the Bethe vectors have fixed
magnetization, therefore these factors multiply to the common
pre-factor in \eqref{OVERLAPS2}.
It is also useful to investigate the diagonal limit of  \eqref{OVERLAPS2}, which is reached by
sending $\beta\to\infty$. In this limit
both $\mathcal{N}(0)$ and $u(\lambda)$ diverge. It can be seen
that the overlap scales to zero unless $N=L/2$, in which case the
previous results are reproduced.

We have tested the conjectured formula numerically and found
convincing agreement in all cases. A short discussion of the numerical
results is presented in the Appendix. In the remainder of the section
 we compute a few specific cases for the overlap.

 The first example is the tilted ferromagnetic state
defined as
\begin{equation}
  \ket{\Psi(F,\theta)}=\prod_{j=1}^L
\frac{1}{\sqrt{2\cosh(\theta)}}
  \begin{pmatrix}
    ie^{\theta/2} \\ e^{-\theta/2}
  \end{pmatrix}.
\end{equation}
Here $\theta=\log(\cot(\vartheta/2))$ with $\vartheta$ being the tilting
angle from the $z$-axis. This state is obtained from \eqref{psyparam} by setting
\begin{equation}
  \alpha=0\qquad \beta=\eta/2+i\pi/2.
\end{equation}
In this case
\begin{equation}
    \mathcal{N}(0)=    -4\sinh^2(\eta)\cosh^2(\theta).
\end{equation}
Thus we get
\begin{equation}
  \label{XFov}
  \frac{|\langle\Psi(F,\theta)|\{\pm\lambda^+\}_{N/2}\rangle|^2}
  {\langle\{\pm\lambda^+\}_{N/2}|\{\pm\lambda^+\}_{N/2}\rangle}
  =
  \frac{|e^{\theta(L-2N)|}}{|2\cosh(\theta)|^{L}}
  \prod_{j=1}^{N/2} u_F(\lambda_j)
  \times \frac{\det_{N/2}G_{jk}^{+}}{\det_{N/2}G_{jk}^{-}}\,,
\end{equation}
with
\begin{equation}
  u_F(\lambda)= \tan^2(\lambda)\tan(\lambda+i\eta/2)\tan(\lambda-i\eta/2).
\end{equation}
It is an interesting idea to consider  the tilted ferromagnetic states
in spin chains with an odd number of sites. In these cases the
integrability properties (or possible off-shell formulas) can not
follow from Boundary Bethe Ansatz methods, because those imply an even
number of sites. However, these one-site states are integrable even in
odd volumes: they are annihilated by all odd charges, which follows
simply from the additivity of the charges and the fact that the states
are integrable in even volumes. Therefore, the overlaps satisfy the
pair structure also in odd volumes. We have tested the formula
\eqref{XFov}  in odd volumes and found that it is indeed correct. This
points to the possibility of a derivation that is independent of the
Boundary BA techniques.

Our second example is the tilted N\'eel state $\ket{\Psi(N,\vartheta)}$ given by the two-site block
\begin{equation}
\ket{\psi}\sim 
  \begin{pmatrix}
-    \cos(\vartheta/2)\sin(\vartheta/2) & \cos^2(\vartheta/2)\\
   - \sin^2(\vartheta/2) &  \cos(\vartheta/2)\sin(\vartheta/2) 
  \end{pmatrix},
\end{equation}
where $\vartheta$ is the tilting angle. This is obtained from
\eqref{psyparam} by setting
\begin{equation}
\alpha=-\eta/2\qquad  e^{-\beta}=\tan(\vartheta/2)\qquad \theta=0.
\end{equation}
Now
\begin{equation}
  \mathcal{N}(0)=-4\sinh^2(\eta) \cosh^2(\beta),
\end{equation}
and for the overlap we get
\begin{equation}
  \frac{|\langle\Psi(N,\vartheta)|\{\pm\lambda^+\}_{N/2}\rangle|^2}
  {\langle\{\pm\lambda^+\}_{N/2}|\{\pm\lambda^+\}_{N/2}\rangle}
  =
  \frac{1}{(2\cosh(\beta))^{L}}
\prod_{j=1}^{N/2}  u_{N,\vartheta}(\lambda_j)
  \times \frac{\det_{N/2}G_{jk}^{+}}{\det_{N/2}G_{jk}^{-}}\,,
\end{equation}
with
\begin{equation}
   u_{N,\vartheta}(\lambda)=\frac{v_{\eta/2}^s(v_{\beta}^c)^2}
  {v^c_{\eta/2}v^s_0v^c_0}.
\end{equation}
It can be seen immediately that in the $\beta\to\infty$ limit the
overlaps with $N=L/2$ converge to those of the original N\'eel state,
whereas for $N<L/2$ they scale to zero.

\section{Conclusions}

In this work we conjectured the formula \eqref{OVERLAPS2}
for exact overlaps with arbitrary two-site states. The
conjecture was confirmed by numerical tests, and we have shown that it
produces the correct Quench Action TBA in the thermodynamic limit.

It is quite intriguing that the new exact formula involves the same
ratio of two Gaudin-like determinants, that were derived originally
for the N\'eel state in \cite{Caux-Neel-overlap1}. For the generalized
dimer states given by \eqref{gendim} this can be understood through
the relation \eqref{Kidentity} which relates the overlaps with them directly to the
N\'eel case. On the other hand, in the generic case we do not know of such a simple
relation which would follow immediately from the coordinate Bethe
Ansatz wave function.

It might be that for generic two-site states it is only the
on-shell case where such a simple final formula can be derived. The
result of \cite{zarembo-neel-cite2} point towards this possibility:
for states corresponding to the special case of lower diagonal
$K$-matrices the off-shell formula differs from that of the N\'eel
state, but the on-shell case already takes the same form. The
recursion relations derived in \cite{calabrese-recursive-overlaps}
also confirm this scenario, because they show that the generic
off-shell case has to have different structure from the original
off-shell formula for N\'eel \eqref{N1}.

In the present
work we have shown that the Gaudin-like determinants play a special role in the
overlap sum rule \eqref{ovsumrule} in the thermodynamic limit: their
ratio tends to a finite number which cancels two other $\ordo(1)$
contributions to the Quench Action. These additional contributions come from the modified
density of states due to the pair structure, and from the fluctuations
around the saddle point solution, in complete analogy with the
$\ordo(1)$ terms in the free energy of a boundary system \cite{sajat-g,woynarovich-uj}.
The cancellation in the $L\to\infty$ limit is necessary for the
consistency of the Quench Action method, but it is not enough to prove
the finite volume formulas.
It is worth noting again, that the same determinants already appeared in a
specific overlap in the Lieb-Liniger model
\cite{caux-stb-LL-BEC-quench,Caux-Neel-overlap2,Brockmann-BEC}, and
analogous determinants were found for overlaps in the
$SU(3)$-symmetric model too \cite{ADSMPS2}. Thus they might be 
generally present in overlaps with integrable initial states.

Future tasks include the construction of a proof of our conjectured
formulas, including the result \eqref{XFov} for one-site states in chains
with odd length. We stress again that the existing methods in the
literature were based on the Boundary Algebraic Bethe Ansatz, and this
only applies to even volumes. Coordinate Bethe
Ansatz calculations don't distinguish between the odd and even cases,
thus they could confirm \eqref{XFov} for low particle
numbers. Alternatively, the recursion relations of
\cite{calabrese-recursive-overlaps} could be adapted to the odd length
case to find a general off-shell formula. 

Also, it would be desirable to develop methods for overlaps in other models. 
In this work we have argued that factorized overlaps can be found only
for integrable initial states. Therefore, in each model the first task
is to characterize the integrable states (by integrable $K$-matrices
or other tools), such that the overlap calculations can be carried out.
An interesting example would be the spin-1 XXZ
chain, where the integrable states are given in 
\cite{sajat-integrable-quenches}, but the exact overlaps are not yet
known. 
An other example
is the $SU(3)$-symmetric chain: the results of \cite{ADSMPS2} for an
integrable MPS are
obtained from coordinate Bethe Ansatz for small particle numbers, and
it is desirable to find a general proof in this case too.

We hope to return to these questions in future research.

\vspace{1cm}
{\bf Acknowledgments} 

We would like to thank M\'arton Mesty\'an, Yunfeng Jiang, Lorenzo Piroli, and Eric
Vernier for inspiring discussions, and we acknowledge support from the ``Premium'' Postdoctoral
Program of the Hungarian Academy of Sciences, the K2016 grant
no. 119204 and the KH-17 grant no. 125567 of the research agency NKFIH. 
This work was partially supported
also  within the Quantum
Technology National Excellence Program (Project No. 2017-1.2.1-NKP-2017-00001).

\appendix

\section{Numerical tests}

We numerically tested the conjectured general formula  \eqref{OVERLAPS2} on finite chains of up to
$L=12$ for various sets of parameters $(\alpha,\beta,\gamma)$. The
methods and computer programs were the same as those described in
\cite{sajat-Corr}: the Bethe states were obtained by diagonalizing
transfer matrices combined with $S_z$, the transfer matrix eigenvalues
were computed numerically for a couple of rapidity parameters, and the
so-called $Q$-polynomial was reconstructed using the $T$-$Q$
relations, which provides and efficient way to determine the Bethe
roots. The overlaps were computed by the real space scalar products,
and compared to the prediction  \eqref{OVERLAPS2}.

Numerical data for the overlap with the tilted ferromagnetic state
$\ket{XF}\equiv \ket{\Psi(F,\pi/2)}$ ($\theta=0$) are listed in Tables
\ref{tab:6}-\ref{tab:12}. The data correspond to $\Delta=2$ and
$L=6,7,8,9,12$.  In the tables we list the energy eigenvalues, the
number of particles, the set of positive rapidities, and the
normalized overlap as obtained from exact diagonalization
(corresponding to the square root of \eqref{XFov}).
We omitted all states that included the singular rapidities
$0,\pi/2$ or exact 2-strings of the form
$\lambda\pm i\eta/2$ with some $\lambda\in\valos$; in these cases
regularized formulas are needed, which are not discussed in the
present work.

The difference
between the prediction and the numerical data for the overlap was
always smaller than $10^{-12}$, therefore we do not list the
errors (at length $L=12$ there were two states with an error of
$\ordo(10^{-5})$, but they had 2-strings very
close to an exact two-string, and this was identified as the source of
the bigger error). We found similar agreement for other $(\alpha,\beta,\theta)$
parameters of the $K$-matrices. We also investigated the $\Delta<1$
regime and found that the overlap formulas are correct for all regular
Bethe states (there are a number of singular cases for $\Delta<1$ \cite{baxter-completeness}, but
these are not discussed here).

\begin{table}
  \centering
  \begin{tabular}{|c|c|l|c|}
    \hline
  $E$ & $N$ & $\{\lambda^+\}_{N/2}$ & $|\skalarszorzat{XF}{\{\pm\lambda\}_{N/2}}|$ \\
    \hline
-22.246211 & 2 & \{0.199867\} & 0.017516245 \\ 
\hline 
-12.000000 & 2 & \{0.785398\} & 0.153093109 \\ 
\hline 
-5.753789 & 2 & \{$\frac{\pi}{2}$+0.705289i\} & 0.458945183 \\ 
\hline 
  \end{tabular}
  \label{tab:6}
  \caption{
    Numerical data for the overlaps with the tilted ferromagnetic state.
 Here $\Delta=2$ and $L=6$.
  }
\end{table}

\begin{table}
  \centering
  \begin{tabular}{|c|c|l|c|}
    \hline
  $E$ & $N$ & $\{\lambda^+\}_{N/2}$ & $|\skalarszorzat{XF}{\{\pm\lambda\}_{N/2}}|$ \\
    \hline
 -22.604374 & 4 & \{$\frac{\pi}{2}$+0.900409i,0.425131\} & 0.054097899 \\ 
\hline 
-19.517541 & 2 & \{0.345713\} & 0.023806863 \\ 
\hline 
-10.904438 & 4 & \{$\frac{\pi}{2}$+1.620317i,1.098201\} & 0.287746749 \\ 
\hline 
-10.610815 & 2 & \{0.917864\} & 0.160905125 \\ 
\hline 
-6.491189 & 4 & \{$\frac{\pi}{2}$+0.668574i,$\frac{\pi}{2}$+2.216935i\} & 0.433258267 \\ 
\hline 
-5.871644 & 2 & \{$\frac{\pi}{2}$+0.683016i\} & 0.370951849 \\ 
    \hline
  \end{tabular}
    \caption{
    Numerical data for the overlaps with the tilted ferromagnetic state.
 Here $\Delta=2$ and $L=7$.
  }
\end{table}

\begin{table}
  \centering
  \begin{tabular}{|c|c|l|c|}
    \hline
  $E$ & $N$ & $\{\lambda^+\}_{N/2}$ & $|\skalarszorzat{XF}{\{\pm\lambda\}_{N/2}}|$ \\
    \hline
-36.157715 & 4 & \{0.676922,0.169306\} & 0.005996275 \\ 
\hline 
-28.296911 & 4 & \{$\frac{\pi}{2}$+0.762821i,0.159158\} & 0.017584753 \\ 
\hline 
-23.126614 & 2 & \{0.137851\} & 0.005629069 \\ 
\hline 
-20.284425 & 4 & \{$\frac{\pi}{2}$+0.785277i,0.581896\} & 0.086678331 \\ 
\hline 
-19.368479 & 4 & \{0.041678,1.336520i\} & 0.000330915 \\ 
\hline 
-17.149343 & 2 & \{0.463317\} & 0.025486372 \\ 
\hline 
-14.529278 & 4 & \{0.564749+0.660013i,0.564749-0.660013i\} & 0.041199177 \\ 
\hline 
-10.614015 & 4 & \{$\frac{\pi}{2}$+1.447586i,1.285445\} & 0.370979758 \\ 
\hline 
-9.790362 & 2 & \{1.019560\} & 0.155431765 \\ 
\hline 
-6.749178 & 4 & \{$\frac{\pi}{2}$+2.041251i,$\frac{\pi}{2}$+0.660208i\} & 0.355324817 \\ 
\hline 
-5.933681 & 2 & \{$\frac{\pi}{2}$+0.671197i\} & 0.290748560 \\ 
\hline 
  \end{tabular}
  \caption{
      Numerical data for the overlaps with the tilted ferromagnetic state.
 Here $\Delta=2$ and $L=8$.
  }
\end{table}

\begin{table}
  \centering
  \begin{tabular}{|c|c|l|c|}
    \hline
  $E$ & $N$ & $\{\lambda^+\}_{N/2}$ & $|\skalarszorzat{XF}{\{\pm\lambda\}_{N/2}}|$ \\
    \hline
-32.137267 & 4 & \{0.297161,0.815441\} & 0.012313623 \\ 
\hline 
-26.460311 & 4 & \{$\frac{\pi}{2}$+0.714053i,0.283467\} & 0.029323961 \\ 
\hline 
-21.449359 & 2 & \{0.246322\} & 0.007529756 \\ 
\hline 
-18.515293 & 4 & \{$\frac{\pi}{2}$+0.730070i,0.715849\} & 0.112794815 \\ 
\hline 
-15.303138 & 2 & \{0.562213\} & 0.024735464 \\ 
\hline 
-13.475088 & 4 & \{0.816328+0.663186i,0.816328-0.663186i\} & 0.079753699 \\ 
\hline 
-10.585428 & 4 & \{$\frac{\pi}{2}$+1.377907i,1.397985\} & 0.389151297 \\ 
\hline 
-9.281427 & 2 & \{1.098337\} & 0.141208039 \\ 
\hline 
-6.826613 & 4 & \{$\frac{\pi}{2}$+1.994065i,$\frac{\pi}{2}$+0.658742i\} & 0.273057092 \\ 
\hline 
-5.966075 & 2 & \{$\frac{\pi}{2}$+0.664996i\} & 0.222944498 \\ 
\hline 
  \end{tabular}
    \caption{
  Numerical data for the overlaps with the tilted ferromagnetic state.
 Here $\Delta=2$ and $L=9$.
  }
\end{table}

\begin{table}
  \small
  \centering
  \begin{tabular}{|c|c|l|c|}
    \hline
  $E$ & $N$ & $\{\lambda^+\}_{N/2}$ & $|\skalarszorzat{XF}{\{\pm\lambda\}_{N/2}}|$ \\
    \hline
-53.840470 & 6 & \{0.841017,0.368953,0.110942\} & 0.000420653 \\ 
\hline 
-48.461935 & 6 & \{$\pi/2$+0.724032i,0.354057,0.107515\} & 0.000815421 \\ 
\hline 
-43.743023 & 4 & \{0.312266,0.097459\} & 0.000213745 \\ 
\hline 
-41.474319 & 6 & \{$\pi/2$+0.744061i,0.749083,0.105807\} & 0.002427526 \\ 
\hline 
-41.396687 & 6 & \{0.229201,0.049088,1.362248i\} & 0.000021161 \\ 
\hline 
-38.224308 & 4 & \{0.599580,0.096548\} & 0.000583217 \\ 
\hline 
-37.591463 & 6 & \{0.669408+0.659265i,0.109511,0.669408-0.659265i\} & 0.001058077 \\ 
\hline 
-37.576158 & 6 & \{$\pi/2$+0.749424i,0.345238,0.742919\} & 0.008821628 \\ 
\hline 
-35.944660 & 6 & \{0.489429,0.024984,1.327009i\} & 0.000014369 \\ 
\hline 
-34.985584 & 4 & \{0.307780,0.596867\} & 0.002112300 \\ 
\hline 
-34.043431 & 6 & \{$\pi/2$+1.428778i,1.344761,0.099967\} & 0.006989255 \\ 
\hline 
-33.462316 & 6 & \{0.617460+0.658894i,0.366883,0.617460-0.658894i\} & 0.003462560 \\ 
\hline 
-32.822751 & 4 & \{1.104307,0.094996\} & 0.002968724 \\ 
\hline 
-30.518925 & 6 & \{$\pi/2$+1.436814i,1.335700,0.321123\} & 0.025203737 \\ 
\hline 
-30.341773 & 6 & \{$\pi/2$+2.029253i,$\pi/2$+0.659301i,0.095578\} & 0.005249464 \\ 
\hline 
-29.657104 & 4 & \{0.302153,1.100640\} & 0.010623914 \\ 
\hline 
-29.551789 & 4 & \{$\pi/2$+0.665839i,0.093343\} & 0.004360074 \\ 
\hline 
-29.325916 & 6 & \{1.320224i,0.979205,0.014355\} & 0.000023370 \\ 
\hline 
-27.116875 & 6 & \{$\pi/2$+2.032170i,$\pi/2$+0.659386i,0.304112\} & 0.018973378 \\ 
\hline 
-26.470339 & 4 & \{$\pi/2$+0.666183i,0.295965\} & 0.015545995 \\ 
\hline 
-26.118545 & 6 & \{0.439354+0.658493i,0.873829,0.439354-0.658493i\} & 0.007751711 \\ 
\hline 
-25.081390 & 6 & \{1.318468i,$\pi/2$+0.684113i,0.009694,\} & 0.000018314 \\ 
\hline 
-24.646110 & 6 & \{$\pi/2$+1.468921i,0.630204,1.303662\} & 0.066154118 \\ 
\hline 
-24.385478 & 4 & \{0.576317,1.090376\} & 0.027193274 \\ 
\hline 
-23.655384 & 2 & \{0.085451\} & 0.000827829 \\ 
\hline 
-21.746288 & 6 & \{$\pi/2$+2.041388i,$\pi/2$+0.659683i,0.579956\} & 0.049754532 \\ 
\hline 
-21.303110 & 4 & \{$\pi/2$+0.667265i,0.559619\} & 0.039052708 \\ 
\hline 
-21.194242 & 6 & \{$\pi/2$+0.700168i,0.344700+0.658480i,0.344700-0.658480i\} & 0.010137832 \\ 
\hline 
-21.064516 & 2 & \{0.267132\} & 0.002871132 \\ 
\hline 
-19.200544 & 4 & \{0.001717,1.317009i\} & 0.000000235 \\ 
\hline 
-18.964761 & 6 & \{1.316961i,0.000448,2.668795i\} & 0.000000015 \\ 
\hline 
-17.389558 & 6 & \{$\pi/2$+1.108583i,1.342005+0.507935i,1.342005-0.507935i\} & 0.103228483 \\ 
\hline 
-16.710970 & 2 & \{0.485898\} & 0.006621650 \\ 
\hline 
-16.200689 & 6 & \{$\pi/2$+2.074803i,$\pi/2$+0.661189i,1.062080\} & 0.215163928 \\ 
\hline 
-15.766469 & 4 & \{$\pi/2$+0.671617i,1.013347\} & 0.141856281 \\ 
\hline 
-15.756503 & 4 & \{0.217016+0.658479i,0.217016-0.658479i\} & 0.001557160 \\ 
\hline 
-14.236805 & 4 & \{0.655902+0.658434i,0.655902-0.658434i\} & 0.012046104 \\ 
\hline 
-13.951915 & 6 & \{0.700809+1.315956i,0.700110,0.700809-1.315956i\} & 0.022711987 \\ 
\hline 
-12.738629 & 6 & \{$\pi/2$+2.087851i,1.384487+0.663893i,1.384487-0.663893i\} & 0.254960253 \\ 
\hline 
-12.385621 & 4 & \{1.185112+0.662562i,1.185112-0.662562i\} & 0.132292301 \\ 
\hline 
-12.000000 & 2 & \{0.785398\} & 0.017116330 \\ 
\hline 
-10.883883 & 6 & \{$\pi/2$+2.686707i,$\pi/2$+1.317465i,1.559164\} & 0.286298034 \\ 
\hline 
-10.654120 & 4 & \{$\pi/2$+1.321734i,1.534546\} & 0.260335581 \\ 
\hline 
-8.573495 & 2 & \{1.248484\} & 0.084054304 \\ 
\hline 
-6.856453 & 4 & \{$\pi/2$+1.975868i,$\pi/2$+0.658480i\} & 0.112477548 \\ 
\hline 
-5.995635 & 2 & \{$\pi/2$+0.659319i\} & 0.093286717 \\ 
\hline 
  \end{tabular}
    \caption{
      Numerical data for the overlaps with the tilted ferromagnetic state.
 Here $\Delta=2$ and $L=12$.
}
  \label{tab:12}
\end{table}

\addcontentsline{toc}{section}{References}
\providecommand{\href}[2]{#2}\begingroup\raggedright\endgroup

\end{document}